\documentclass[twocolumn]{article}
\usepackage{epsfig,deluxe}
\begin{document}

\author{A. De R\'ujula }
\title{\bf{Optical and X-ray Afterglows in the  Cannonball Model of GRBs}}

\author{ A. De R\'ujula}

\maketitle           

\begin{abstract}

The Cannonball Model is based on the hypothesis that GRBs
and their afterglows are made in supernova explosions by
relativistic ejecta similar to the ones observed in quasars
and microquasars. Its predictions are simple, and analytical
in fair approximations. The model describes well the 
properties of the $\gamma$-rays of GRBs. It gives a
very simple and extremely successful description of the
optical and X-ray afterglows of {\it all} GRBs of known
redshift. The only problem the model has, so far, is
that it is contrary to staunch orthodox beliefs.

\end{abstract}

\date{\today}
\maketitle

\section{Introduction}

The idea that GRBs are due to collimated emissions is not recent.
In the case of GRBs from quasars it was discussed by Brainerd \cite{Bra}; 
 in the case of a funnel in an explosion, by Meszaros \&
Rees \cite{MR}. In what is no doubt the relevant case: jets in 
stellar gravitational collapses, the idea has been developed
over the years by Dar and collaborators 
\cite{SD} to \cite{DD2001c}.
Now we know that long-duration GRBs are cosmological,
originate in galaxies, are associated with supernovae (SNe) and have
energies that would be ridiculously large for a stellar spherical explosion
(a fireball). GRBs must be ``jetted''.

In the currently dominating scenarios, the ejecta that beget a GRB and its
afterglow (AG) are thrown off in a uniform cone with an opening angle
$\theta_j(0)$. This cone expands sideways: the angle $\theta_j(t)$
subtended by the ejecta {\it increases} with time, delineating a {\it firetrumpet}, 
as in Fig.~(1). Relativistic jets are ubiquitous in astrophysics.
The ejecta of these real jets, as seen from their emission
point up to the point where they eventually stop and expand, generally
subtend angles that {\it decrease} with time: just the opposite
of the assumed behaviour of firetrumpets.
In the analysis of the observed jets, e.g. \cite{PZ,RM,GC},
it is the fixed angle of observation ---and not the angle subtended by
the ejecta--- that plays a key role.

\begin{figure}
  \includegraphics[height=.5\textheight]{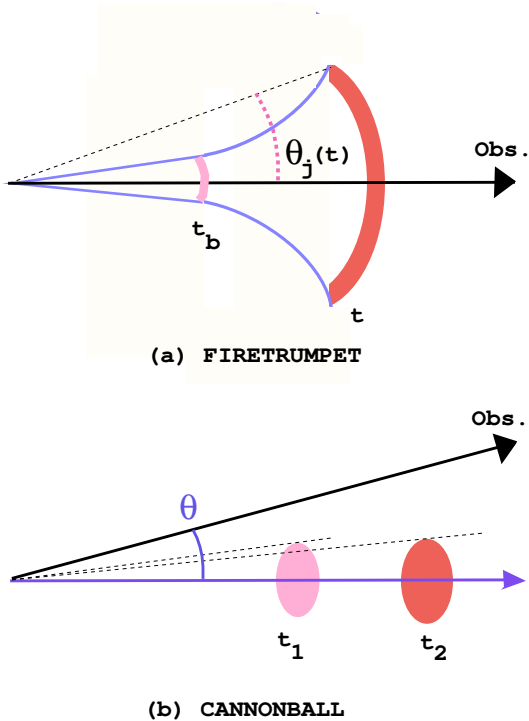}
\caption{(a) A firecone or, more properly, a {\it firetrumpet}.
In these scenarios the cone expands
conically for a distance, after which the jet angle
$\rm\theta_j$ widens faster as its firefront travels.
(b) Cannonballs (shown here, somewhat pedantically,
a bit Lorentz-contracted) subtend decreasing angles as they travel.
The only relevant angle in the CB model is the observer's
viewing angle $\theta$.}
\end{figure}

The Cannonball (CB) model is based on the contention that
GRBs and their AGs are made by relativistically jetted balls of ordinary 
``baryonic'' matter
which, by a mechanism \cite{DD2001c} that I will outline,
stop expanding soon after their emission. 
The CB idea gives a good description of the properties of the $\gamma$-rays
in a GRB, that we modelled in simple approximations \cite{DD2000b}. 
It gives an excellent and complete description of optical and
X-ray afterglows, which we have modelled in full detail, as I outline here.

\section{The GRB and its engine}

We assume that in  core-collapse SN events
a tiny fraction of the parent star's
material, external to the newly-born compact object,
falls back in a time of the
order of very roughly one day \cite{DeR87, DD2000a}.
Given the considerable
specific angular momentum of stars, it settles into an accretion disk 
around the compact object. The subsequent sudden episodes
of catastrophic
accretion ---occurring with a chaotic
time sequence that we cannot predict---
result in the emission of CBs, lasting till the reservoir of accreting matter is 
exhausted. The emitted CBs initially expand in their rest system at a speed
$\beta_T\,c$, of the order of
the speed of sound in a relativistic plasma ($\beta_T=1/\sqrt{3}$).
These considerations are illustrated in Fig.~(2).

\begin{figure}
  \includegraphics[height=.7\textheight]{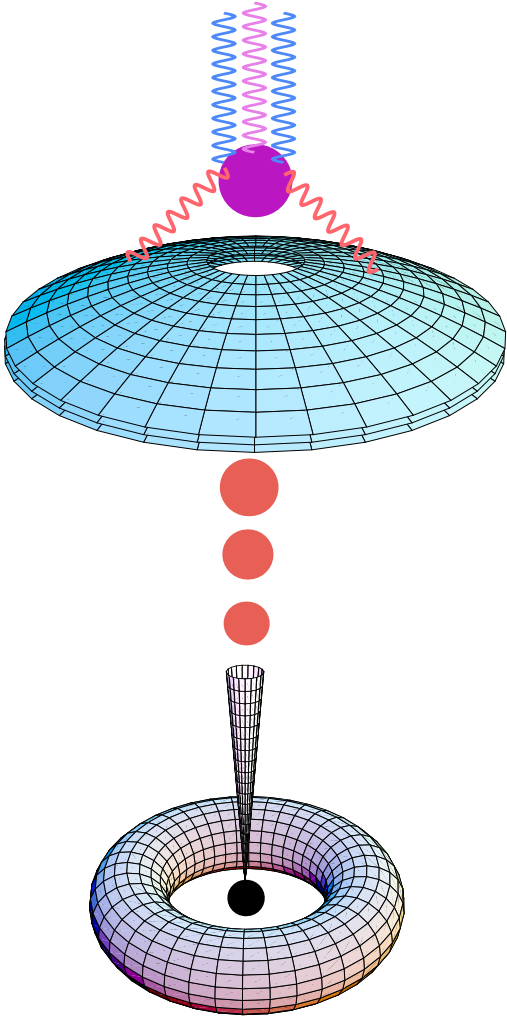}
\caption{An ``artist's view'' (not to scale) of the CB model
of GRBs and their AGs. A core-collapse SN results in
a compact object and a fast-rotating torus of non-ejected
fallen-back material. Matter (not shown) catastrophically accreting
into the central object produces
a narrowly collimated beam of CBs, of which only some of
the ``northern'' ones are depicted. As these CBs pierce the SN shell,
not precisely on the same tiny spot,
they heat and re-emit photons, which are
Lorentz-boosted and collimated by the CBs' relativistic motion.}
\end{figure}

From this point onwards, the CB model relies on processes whose outcome 
can be approximately worked out in an explicit manner. The collision of
the CB with the SN shell heats the CB (which is not transparent at this point
to $\gamma$'s from $\pi^0$ decays) to a surface temperature that, by the
time the CB reaches the transparent outskirts of the SN shell, is
$\sim 150$ eV, further decreasing as the CB travels \cite{DD2000b}. 
The resulting quasi-thermal CB surface
radiation, Doppler-shifted in energy and forward-collimated by the CB's
fast motion, gives rise to an individual pulse in a GRB \cite{DD2000b}. The GRB
light curve is an ensemble of such pulses, often overlapping one another.
The energies of the individual GRB $\gamma$-rays, as well as their typical
total fluences, require CB Lorentz factors $\gamma$ of ${\cal{O}}$(10$^3$).

The CB model also explains the ``Fe lines'' seen in some X-ray AGs,
as boosted hydrogen-recombination lines \cite{DD2001a}. Their properties require
$\gamma\sim 10^3$ and a baryonic number per cannonball 
$N_{CB}\sim 6\times 10^{50}$. Even in a GRB with very many significant pulses,
the total mass of a jet of CBs would be comparable to that of the Earth: peanuts,
by stellar standards.

The rest of this note is based on
\cite{DD2001c}, and concentrates on afterglows, for which the CB theory
is very simple.

\section{Optical Afterglows: theory}

When an expanding CB, in a matter of (observer's) seconds, becomes transparent 
to its enclosed radiation, it loses its internal radiation pressure. If it has been
expanding  at a speed comparable to that of relativistic
sound, should it not inertially continue to do so? No! 
We assume CBs to enclose a magnetic field maze, as the
observed ejections from quasars and microquasars do.
The interstellar medium (ISM) the CBs traverse has been previously partially
ionized by the forward-beamed GRB radiation. The neutral ISM fraction
is efficiently ionized by Coulomb interactions as it enters the CB.
In analogy to processes occurring in quasar and microquasar
ejections, the ionized ISM particles are multiply
scattered, in a ``collisionless'' way, by the CBs' turbulent magnetic fields.
In the rest system of the CB the ISM swept-up nuclei are isotropically
re-emitted, exerting
an inwards force on the CB's surface. This allows one to compute {\it explicitly}
the CB's radius as a function of time \cite{DD2001c}.
The radius, for typical parameters, reaches 
a constant value $R_{max}\sim 10^{14}$ cm in minutes of observer's time.
 
The ISM nuclei (mainly protons) that a CB scatters also decelerate
its flight: its Lorentz factor, $\gamma(t)$, is calculable.
Travelling at a large $\gamma$ and viewed at a small angle
$\theta$, the CB's emissions are strongly relativistically aberrant:
in minutes of observer's time, the CBs are parsecs away from their source.
For a constant CB radius and an approximately constant ISM density,
$\gamma(t)$ has an explicit analytical expression \cite{DD2001c}. Typically 
$\gamma=\gamma(0)/2$ at a distance of order 1 kpc from the source.

The ISM electrons entering a CB bounce off
its enclosed magnetic domains, lose energy effectively
by synchrotron radiation, and acquire a predictable power-law energy
spectrum, $dn_e/dE\propto E^{-3.2}$, which implies a given distribution of the 
radiated photons: $\nu\,dn_\gamma/d\nu\propto \nu^{-1.1}$ \cite{DD2001}.
The emitted energy rate, in a CB's rest system, is equal to the rate  at which 
the ISM electrons bring energy into the CB\footnote{The 
kinetic energy of a CB is mainly lost to the ISM protons it scatters; only a fraction 
$\sim m_e/m_p$ is re-emitted by electrons, as the AG.}. The AG fluence in the CB
model is of the form:
\begin{equation}
F_\nu=f\,\nu^{-\alpha}\;[\gamma(t)]^{2\alpha}\,
\left[{2\,\gamma(t)\over 1 + \gamma(t)^2\,\theta^2}\right]^{3+\alpha},
\label{fluence}
\end{equation}
with $\alpha\sim 1.1$ and
$f$ an {\it explicit} normalization proportional to the ISM electron density,
$f\propto n_e$.

We assume that all long-duration GRBs are associated with SNe and,
as a bold ansatz, we take these SNe to have the fluence of SN1998bw,
properly transported in time and frequency to the GRB's redshift
\cite{D99b}.
An observed AG's fluence is then the sum of this SN, the background 
galaxy emission, and the CBs' contribution, Eq.~(\ref{fluence}).
Remarkably, this very simple theory very successfully describes the optical
AGs, {\it at all times}, of {\it every} GRB of known redshift.

\section{X-ray afterglows: theory}

In the CB model, the
X-ray emission by a GRB is more complex than its
optical emission.
During the GRB the emitted light at all energies is mainly of
thermal origin (although it does not have a thermal spectrum)
and, in a fixed energy interval, it decreases exponentially
with time \cite{DD2000b}. A few seconds after the last GRB pulse
(the last CB), this pseudothermal emission becomes a
subdominant effect. For the next few hours, the evolution
of a CB is interestingly complicated. In particular, its originally
ionized material should recombine into  hydrogen and
emit Lyman-$\alpha$ lines that are seen Doppler-boosted
to keV energies \cite{DD2001a}. Later, the
CBs settle down to a much simpler phase, which typically lasts
for months, till the CBs finally stop moving relativistically.

The X-ray AG is initially dominated by thermal
bremsstrahlung (TB), which has a harder spectrum than 
synchrotron emission. This period of TB-dominance
begins at a time $t_{trans}$,
a few seconds after the end of the GRB, when the last CB
becomes transparent to its enclosed radiation.
A few minutes later,  both the
X-ray and optical AGs are dominated by synchrotron
radiation, and their shapes are {\it achromatic}, as in
Eq.~(\ref{fluence}).

The TB X-ray fluence decreases with time as $R^{-3}\,T^{1/2}$,
with $R$ the CB's radius, still increasing linearly with time at an
early stage. Depending on whether a CB's cooling soon after 
$t_{trans}$ is dominated by TB, or by adiabatic losses, the
X-ray fluence is $\propto t^{-5}$ or $\propto t^{-4}$, the first behaviour
being expected for our typical CB parameters.

The previous considerations justify a very simple
description of the X-ray light curves:
\begin{equation}
F_X(t) \simeq f_X(t_{trans})\, \left[{t_{trans}\over t}\right]^5
+ F_{sync}(t) \, ,
\label{Xdensity}
\end{equation}
where $ t$ is the observer's time since the ejection of the (last) CB, and
 $F_{sync}(t)$ is the synchrotron fluence in the X-band,
i.e. Eq.~(\ref{fluence}) integrated in the relevant energy interval. 
The normalization $f_X$ is, once again, explicit in the CB model.
Equation (\ref{Xdensity}) provides an excellent description,
{\it at all times,} of {\it all} the
X-ray AGs of GRBs of known $z$.

\section{Optical afterglows: results}

GRBs have varied numbers, $n_{CB}$, of gamma-ray pulses, or CBs,
which may have different initial Lorentz factors $\gamma_0$ and 
baryon numbers $N_{CB}$. This and other complications are
eased by the fact that the AG light curve is the sum of temporally 
unresolved individual CB afterglows: we can characterize,
as in Eq.~(\ref{fluence}), the
AG with the parameters of one single CB, whose values
represent a weighted average.
The parameters to be fit are $f$, $\theta$ and $\alpha$ 
in Eq.~(\ref{fluence}), as well as two parameters entering the
expression for $\gamma(t)$: $\gamma_0$ and the deceleration
parameter $x_\infty\equiv N_{CB}/(\pi\, R_{max}^2\, n_p)$, with
$n_p$ the ISM proton number density
light-centuries away from the GRB's progenitor.

The fits to the CB model are generally excellent.
An example, GRB 970228, is shown in Fig.~(3). Our only bad fit,
that to GRB 000301c, is shown in Fig.~(4). The occasional
misfit is to be expected: the AG fluences
are proportional to $n_e$, which is
not constant for kpc distances, even
in the halo of galaxies. The fitting procedure
---which attributes to the errors a counterfactual purely-statistical origin---
results in tiny 1 $\sigma$ spreads for the parameters, and in
excellent confidence levels, on which one {\bf should not} place 
excessive confidence.

\begin{figure}
 \includegraphics[height=.3\textheight]{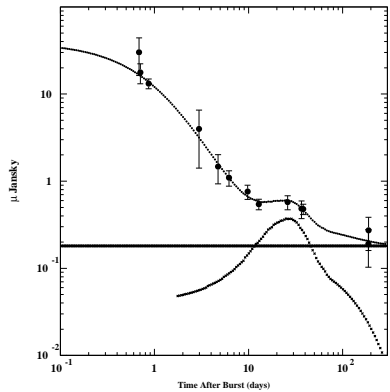}
\caption{Comparison between the fitted R-band afterglow
(upper curves) and the observations
for GRB 970228, at $z=0.695$, 
without subtraction of the host
galaxy's contribution (the straight line).
The contribution
from a 1998bw-like supernova placed at the GRB's
redshift
is indicated by a line of crosses.
The SN bump is clearly discernible.}
\end{figure}

\begin{figure}
  \includegraphics[height=.3\textheight]{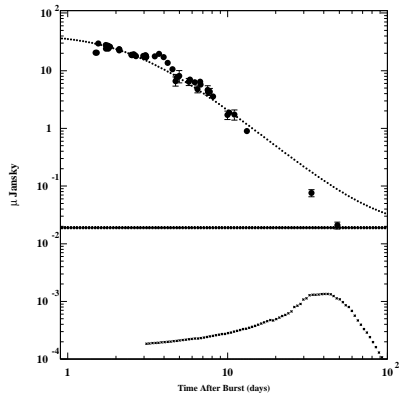}
\caption{Comparison between the fitted R-band afterglow
and the observations
for GRB 000301c, at $ z=2.040$, 
without subtraction of the host
galaxy's contribution (the straight line).
The contribution
from a 1998bw-like supernova placed at the GRB's
redshift is too weak to be
observable.}
\end{figure}

The spectral slope in the optical-to-X-ray interval, $\alpha$, is
the only parameter for which we have no reason to expect
a range of different values. It is extremely
satisfactory  that the fitted values of $\alpha$ 
(extracted from the AG's temporal shape at fixed $\nu$) are, within errors,
compatible with {\it all} of the GRBs having a universal
behaviour with the theoretically predicted value: $\alpha\approx 1.1$.
Most spectral measurements agree with this result and, for the ones that
do not (and some of the ones that do) there is always a good reason not to worry:
poorly-understood absorption.
The distribution of $\gamma_0$ values agrees snugly with our expectation
from the properties of the GRB: $\gamma_0\sim 10^3$, and
it is surprisingly narrow: $\Delta \gamma_0/\gamma_0\sim 0.2$.
The values of $x_\infty$ should be fairly spread: they depend on $n_p$
close to the progenitor (that determines $R_{max}$) and $n_e$ in the region
where the AG is emitted. Indeed, the $x_\infty$ values range for an
order of magnitude above and below a typical prediction.
The overall normalization $f$ of the optical afterglows is, with the 
other parameters fixed, $f\propto n_{CB}\,N_{CB}$. The results from
optical AG  fits range for an order of magnitude above and below
the prediction for a single dominant CB, $n_{CB}=1$, and our typical
expected $N_{CB}$. This must be partially due to absorption
uncertainties, for the X-ray fits result in 1/3 as much spread.

To summarize, the distributions of {\it all} parameters are in extremely good
agreement with the expectations of the CB model and, if anything,
they are astonishingly close to what they would be for ``standard candle'' GRBs.

\subsubsection{GRB 970508: a case of gravitational lensing}

The AG of GRB 970508 has a most peculiar shape. An attempt to fit it
with Eq.~(\ref{fluence}) is shown in Fig.~(5): it is a miserable failure.
But suppose the light from the CBs of this GRB is gravitationally
lensed by a star or a binary, of mass $\sim 2\,M_\odot$, placed
roughly half-way to their position (the probability for something like this to
happen is a few per-cent). The lensed AG, whose CB model parameters
are entirely conventional, is shown in Fig.~(6): the fit is fantastically
good. Comparing Figs.~(5) and (6), notice how time-asymmetric
the amplification is (the time scale is logarithmic!). This is because,
as the lensing occurs, the CBs are --- in a specific predicted fashion---
slowing down from an initial superluminal speed $\rm v_\perp\sim 500$ c.
Seing a result like this, one knows {\it in one's bones} that one is on the
right track!

\begin{figure}
  \includegraphics[height=.35\textheight]{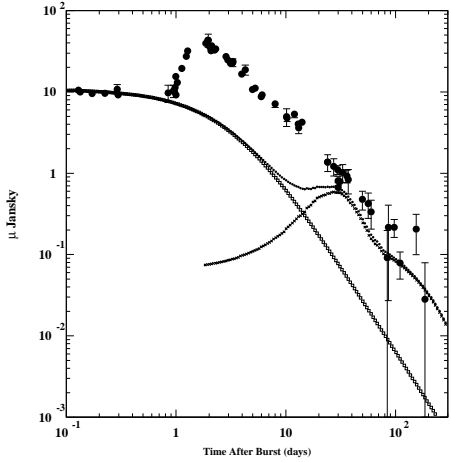}
\caption{Comparisons between a fitted R-band afterglow
(upper curves) and the observations,
for GRB 970508, at $z=0.835$. 
The galaxy has been subtracted. The fit is a total disaster.}
\end{figure}

\begin{figure}
  \includegraphics[height=.35\textheight]{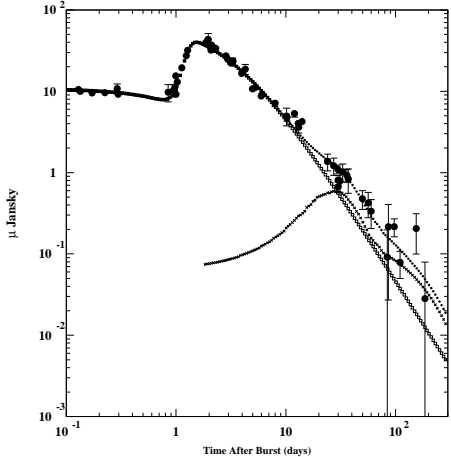}
\caption{The R-band AG of GRB 970508, fit with the additional
effect of gravitational lensing by a $\sim 2\,M_\odot$ intervening
object.
This time the fit is excellent.
A SN1998bw-like contribution is necessary. }
\end{figure}

\subsubsection{GRB 990123 and its mother's wind}

For this GRB, there are good optical data  starting
exceptionally early: $22.18$ seconds after 
its detected beginning \cite{Ak}. 
The AG rises abruptly to a second point at $t=47.38$ s,
 and decreases thereafter. We can very simply describe
this AG from this point onwards, as shown in Fig.~(7).
From the fitted values of $\rm z$, $\gamma_0$ and $\theta$,
we  conclude that at that time the CBs are a mere $x=0.46$ pc away
from the progenitor star. This is precisely where the 
density profile ought to be  $n\propto r^{-2}$, induced
by the parent-star's wind and ejecta. 
This early, the CB's deceleration is negligible: an
$r^{-2}$ density profile implies an
optical AG declining as $t^{-2}$. The shape
and normalization of the early AG are precisely
the expected ones (the inferred local density is
$\rm 0.54 \,cm^{-3}$ at $\rm x\simeq 0.46$ pc).
The CB model describes the full history of an
optical AG.

\begin{figure}
  \includegraphics[height=.33\textheight]{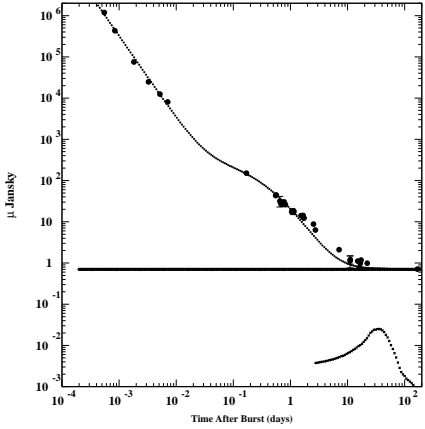}
\caption{Comparisons between the fitted R-band AG
and the observations,
 for GRB 990123, at $z=1.600$,
without subtraction of the host galaxy.
The data on this AG begins shortly after the GRB. 
The starting $t^{-2}$ behaviour is that expected if the CBs are moving trough a
density profile, $n\propto r^{-2}$,
induced by the parent-star's pre-SN wind and
ejections.}
\end{figure}

\section{X-ray afterglows: results}

Once again, I can only show a subsubset of our results.
The X-ray AG of GRB 010222, and its CB-model's description,
are shown in Fig.~(8). The early decline is dominated by thermal
bremsstrahlung and has the expected $t^{-5}$ decline. The 
late AG is achromatic: the shape of the late X-ray fluence is that 
obtained with the parameters resulting from the fit to the 
optical AG. The normalizations at early and late
times are in the expected range.  The approximation of a constant ISM
density in the normalization of $F_{sync}(t)$ in Eq.~(\ref{Xdensity})
should be inappropriate between
$ \rm \sim\! 2\times 10^{-3}$ and $\sim\! 0.2$ days. There are no
data in that domain except for GRB 991216 and perhaps 970508, which
 suggest an initial density variation 
$\rm\propto 1/r^2$, resulting in an observed $\rm\sim t^{-2}$ decline,
as in the optical AG of GRB 990123, shown in Fig.~(7).

Our worst but most significant
fit to an X-ray AG is that to GRB 980425, shown in Fig.~(9).
Unlike the observers \cite{Elena} we assume the AG
was produced by the CBs and {\it not} by conventional
ejecta of the associated SN 1998bw, from which a much weaker 
signal is to be expected.

\begin{figure}
  \includegraphics[height=.33\textheight]{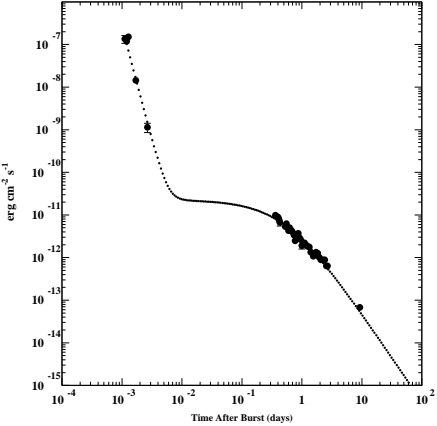}
\caption{The early-time and late-time X-ray AG of GRB 010222 in the 2--10
keV, fitted with a constant density along the 
CB trajectory. The early decline is $\propto t^{-5}$; the late behaviour
is achromatic: ``parallel'' to the optical AG curve.}
\end{figure}

\subsection{Even GRB 980425 is ``normal''}

The optical AG of this close-by GRB is dominated by 
SN1998bw, and we have extracted
its AG parameters ($\theta$, $\gamma_0$, etc.) 
from the X-ray fit of Fig.~(9).
The fitted viewing angle is $\theta\sim 8.3$ mrad;
$\gamma_0\sim 750$, etc. are normal.
If the CBs of GRB 980425 had been viewed from a typical viewing
angle, $\theta\!\leq\! 1/\gamma_0 $, the equivalent isotropic energy
would have been in the range of all other GRBs.
If for this case the ISM density and CB
radius were the same as for other GRBs, 
the predicted intensity of the X-ray 
``plateau'' is
$\rm\!\sim\! 4\times 10^{-13}\, erg\, cm^{-2}\, s^{-1}$, 
as observed \cite{Pian}.
The ``normal''  GRB energy and X-ray AG fluence strongly support 
the association 
of SN1998bw with (a not exceptional) GRB 980425.

\begin{figure}
  \includegraphics[height=.35\textheight]{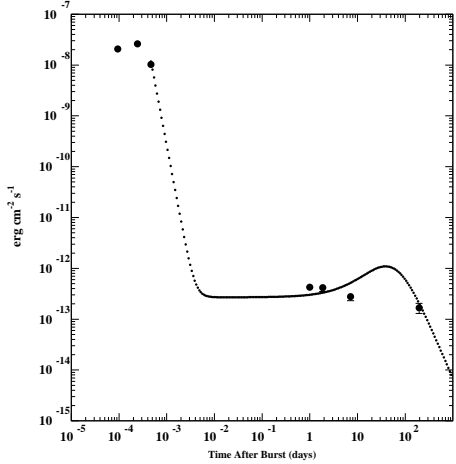}
\caption{CB-model fit to the X-ray afterglow of the SN1998bw/GRB 980425
pair. The flatish domain we call ``plateau''. It is so extensive because
$\theta\,\gamma_0\gg 1$ and it takes time to reach the maximum at
$\theta\,\gamma(t)\sim 1$.}
\end{figure}

The  parameters of the X-ray AG can be used to predict the
magnitude and shape of the
optical AG of the blended SN 1998bw/GRB 980425 system, see
Fig.~(10). The CBs' contribution dominates at late time and is
in perfect agreement with the HST
observation \cite{Fy} on day 778. At that time the SN and the
CB (this is a single-pulse GRB) were far enough from each other,
and close enough to us, to be resolvable! \cite{DD2000a}.
Alas, the rare occasion was missed.

Interpreted in the CB model, 
the fluence and soft spectrum of GRB980425, as well as its X-ray and 
optical AGs were ``normal''. It was simply much closer, and viewed
at a much larger angle than other GRBs.

\begin{figure}
  \includegraphics[height=.25\textheight]{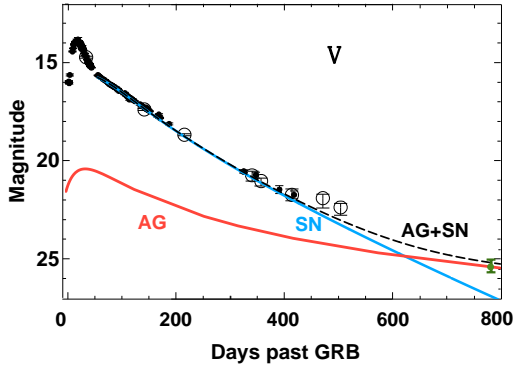}
\caption{The V-band light curve of SN1998bw/GRB 980425.
The blue ``SN'' curve is a fit to the SN \cite{So},
dominated after day $\sim\!40$ by $^{56}$Co decay.
 The red ``AG'' curve is our prediction for the CB-induced
AG component, as given by Eq.~(\ref{fluence}),
with the parameters determined from the X-ray AG fit in the
previous figure. The SN contribution
dominates up to day $\sim\! 600$. The last point is an
HST measurement at day 778, that precisely agrees with the (dashed)
SN plus CB prediction for the total AG.}
\end{figure}

\begin{figure}
  \includegraphics[height=.35\textheight]{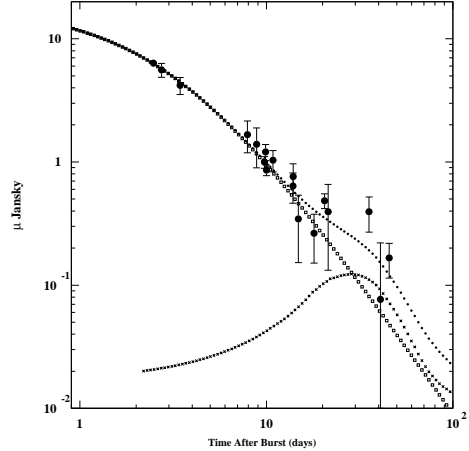}
\caption{
Comparisons between our fitted R-band afterglow
 and the observations for GRB 000418, at $\rm z=1.119$. The galaxy has been subtracted. The contribution
from a 1998bw-like SN placed at the GRB's redshift
is indicated by a line of crosses.  There is a significant indication of a
SN1998bw-like contribution.}
\label{fig418}
\end{figure}

\begin{figure}
  \includegraphics[height=.35\textheight]{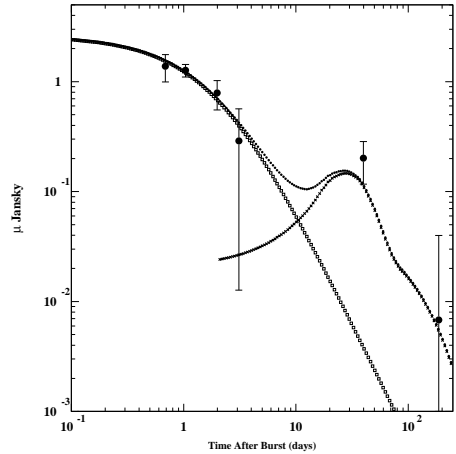}
\caption{The same as Fig.~(\ref{fig418}), but for GRB 980613, at $\rm z=1.096$. 
 A SN 1998bw-like contribution, 
though based on
only one significant point, appears to be required.}
\end{figure}

\begin{figure}
 \includegraphics[height=.35\textheight]{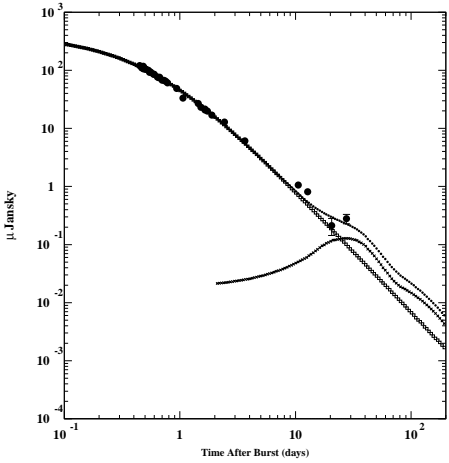}
\caption{The same as Fig.~(\ref{fig418}), but for
 GRB 991216, at $\rm z=1.020$. The data show possible 
evidence for a SN bump.}
\end{figure}

\begin{figure}
 \includegraphics[height=.35\textheight]{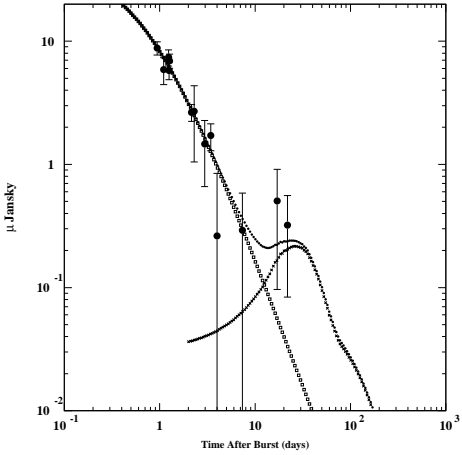}
\caption{The same as Fig.~(\ref{fig418}), but for GRB 980703, at $\rm z=0.966$. A SN 1998bw-like contribution, though the errors
are large, appears to be required.}
\end{figure}

\begin{figure}
\includegraphics[height=.35\textheight]{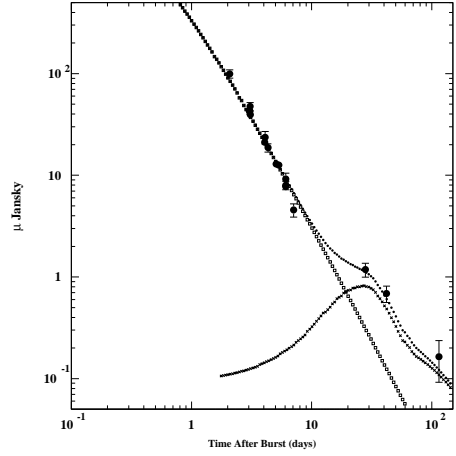}
\caption{The same as Fig.~(\ref{fig418}), but for GRB 991208, at $\rm z=0.706$. The SN contribution is clearly discernible.}
\end{figure}

\begin{figure}
 \includegraphics[height=.35\textheight]{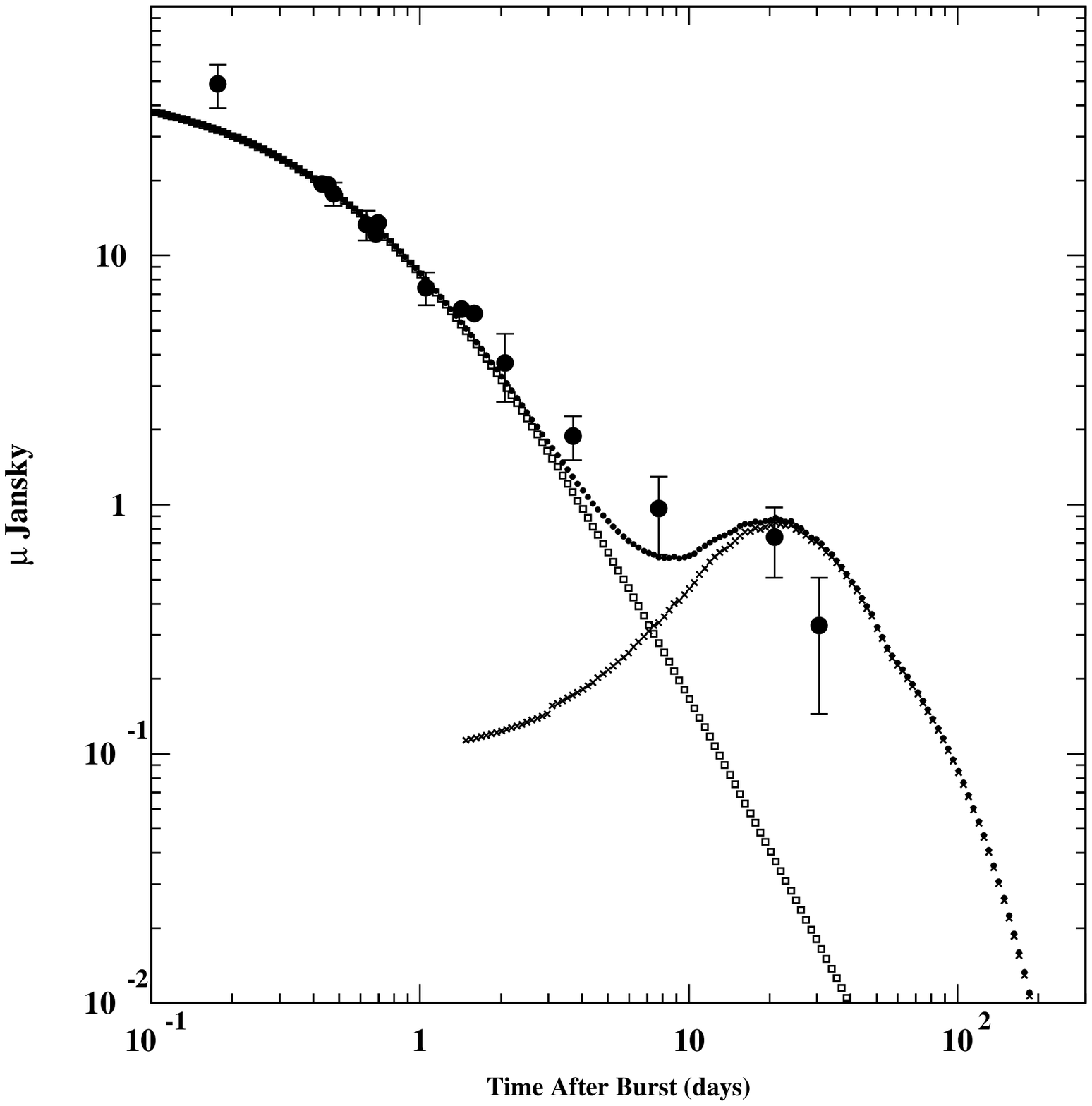}
\caption{The same as Fig.~(\ref{fig418}), but for GRB 990712, at $\rm z=0.434$. The SN contribution is clearly discernible, but a bump at slightly
earlier times than that of our standard-candle
SN1998bw would provide a slightly better description.}
\end{figure}

\subsection{Are GRBs associated with SNe?}

The complete success of the CB model in describing optical AGs
results in an excellent exposition \cite{DD2001c} of the GRB/SN association.
It is useful to discuss the issue in order of diminishing redshift.
In the six more distant GRBs there is
no evidence for {\it or against} a SN 1998bw-like component,
two examples are given in Figs.~(4) and (7).
In the next five closer cases there is evidence, ranging from
fairly weak to very strong; see Fig.~(6) for GRB 970508,
for which the light curve ---in the CB model--- clearly
requires a SN component. The other seven of these cases
---000418, 980613, 991216, 980703, 991208, and 990712, shown in 
Figs.~(11) to (16) and 970228, shown in Fig.(3)--- a SN1998bw-like 
contribution is  required, to varying levels of statistical
confidence. Finally, GRB 980425, at $z=0.0085$ is 
indeed associated to a SN \cite{Ga}.
The general trend is clear. With its excellent description of afterglows,
the CB model's predictions for the ``proper'' AG from the CBs  
provide an excellent ``background'' on top of which to discern a SN
contribution. The conclusion is that, the closer a GRB is, the better 
the evidence for a SN. For the more distant GRBs,
a SN could not be seen, even if it was there.
In all cases where the SN could be seen, it was seen; the evidence
gaining in significance as the distance diminishes.
The conclusion that all long-duration GRBs are
associated with SNe is irresistible.

\subsection{What fraction of SNe emit GRBs?}

The $\gamma$-ray fluence of a GRB \cite{DD2000a} is
$\propto [2\,\gamma_0/(1+\gamma_0^2\,\theta^2)]^3$.
The $\theta$ dependence is the steepest
parameter dependence of the CB model. It is 
reasonable to attribute the range of equivalent spherical energies
mainly to the $\theta$ dependence (as if GRBs were otherwise
standard candles). Excepting GRB 980425, the observed  spread in
equivalent energy then corresponds to a spread 
$\theta \approx 0$ to $\theta_{max}\approx 2.4/\gamma$.
Thus the fraction of observable GRBs (with the
current or past sensitivity) is $f(\gamma)=2\pi\, \theta_{max}^2 / (4
\pi) \approx 2.84 / \gamma^2 $, with two jets of
CBs per event. The rate of Type II/Ib/Ic SNe in the observable universe,
multiplied by $f(10^3)$, coincides with
the observed rate of GRBs, as if every SN emitted GRBs!
But the observational
errors are {\it very} large; distant SNe may be more frequent
than in current estimates, if
the star formation rate continues to increase above $z\sim 1$;
the efficiency of GRB detection as a function of fluence has not
been taken  into account in this estimate.  In spite of these uncertainties,
the conclusion, in the CB model, is that {\it a very
large fraction of SNe  emit GRBs.}
Then, why is SN1998bw peculiar? We have seen that
in X-rays it is not: they were CB-induced. Other
peculiarities should also be due to how close to
the GRB ``axis'' this SN was observed.

\section{Conclusions and social affairs}

At the time I gave this talk and submitted it to the Woods Hole GRB
conference proceedings, we had not yet worked out the CB-model's 
predictions for radio afterglows. By the time I am posting it on the web,
we have made significant progress: once again, the CB model's predictions
are uncharacteristically simple and succesful \cite{DDD2002}.
We have no CB-model explanation for the
scintillation behaviour of GRB 970508 \cite{Ta}. Other than that, the CB model
explains well all properties of GRBs. In the case of optical
and X-ray afterglows, which I have outlined,
the CB-model's predictions 
are univocal (as opposed to multiple-choice), very explicit,
analytical in fair approximations, quite simple,
very complete,  and extremely successful.
I doubt that this statement applies to other models of GRBs.

I am not saying that the CB model will stand all future tests. But I
would expect that ---confronted with a simple and successful model---
most scientists would, at least, say: {\it Hum! } and ask
good questions. Not the case. Four of
our papers on this subject, \cite{DD2000a} to \cite{DD2001b},
have already been rejected by referees who found no single error,
and/or stuck to bad numerical questions, even after being proved 
numerically (i.e. inarguably) wrong. This rejection statistics
makes me feel that the GRB community has 
officially certified us... as crackpots. All this reflects, I suspect,
the global rise of fundamentalism. Some people,
almost literally in this case, still refuse to ``look through the telescope''.
A GRB theorist, sitting on the first row in my talk, was kind enough
to enact my social comments. Indeed, he rose in ire to
exclaim: {\it I am glad that the referee system is working!} He then
stuck to a bad numerical question. Most of the rest of the questions
period was not this aggressive\footnote{Most... but not all.
Another GRB theorist stated his suspicion that I was not presenting
radio results, not because we had not yet worked them out, but because
they were no good!!! For my first experience at a GRB conference,
this was not bad: a bit like dancing flamenco on a mine field.}: there 
is still hope in science's sempiternal contest with faith.

\end{document}